\documentclass[krantz1,ChapterTOCs]{krantz} 
\usepackage{fixltx2e,fix-cm}
\usepackage{amssymb}
\usepackage{amsmath}
\usepackage{graphicx}
\usepackage{subfigure}
\usepackage{makeidx}
\usepackage{multicol}

\makeindex

\begin{document}

\newtheorem{theorem}{Theorem}

\frontmatter

\title{Regression approaches for Approximate Bayesian Computation} 
\author{Michael G.B. Blum, CNRS Universit\'e Grenoble Alpes}
\maketitle

\cleardoublepage
\setcounter{page}{7} 

\mainmatter

\section{Content}\label{content}
This chapter introduces regression approaches and regression adjustment for Approximate Bayesian Computation (ABC). Regression adjustment adjusts parameter values after rejection sampling in order to account for the imperfect match between simulations and observations. Imperfect match between simulations and observations can be more pronounced when there are many summary statistics, a phenomenon coined as the {\it curse of dimensionality} \cite{blum10}. Because of this imperfect match, credibility intervals  obtained with regression approaches can be inflated compared to true credibility intervals \cite{csillery10}. The chapter presents the main concepts underlying regression adjustment. A theorem that compares theoretical properties of posterior distributions obtained with and without regression adjustment is presented. Last, a practical application of regression adjustment in population genetics shows that regression adjustment shrinks posterior distributions compared to rejection approaches, which is a solution to avoid inflated credibility intervals.

\section{Introduction}\label{intro}

In this chapter, we present regression approaches for Approximate Bayesian Computation (ABC). As for most methodological developments related to ABC, regression approaches originate with coalescent modeling in population genetics \cite{beaumont02}. After performing rejection sampling by accepting parameters that generate summary statistics close enough to those observed, parameters are {\it adjusted} to account for the discrepancy between simulated and observed summary statistics. Because adjustment is based on a regression model, such approaches are coined as {\it regression adjustment} in the following.

Regression adjustment is a peculiar approach in the landscape of Bayesian approaches where sampling techniques are usually proposed to account for mismatches between simulations and observations \cite{marjoram03,sisson07}. We suggest various reasons explaining why regression adjustment is now a common step in practical applications of ABC. First, it is convenient and generic because the simulation mechanism is used to generate simulated summary statistics as a first step and it is not used afterwards. For instance, the software {\it ms} is used to generate DNA sequences or genotypes when performing ABC inference in population genetics \cite{hudson02}. Statistical routines, which account for mismatches, are completely separated from the simulation mechanism and are used in a second step. Regression adjustment can therefore be readily applied in a wide range of contexts without implementation efforts. By contrast, when considering sampling techniques, statistical operations and simulations are embedded within a single algorithm \cite{sisson07,beaumont09}, which may require new algorithmic development for each specific statistical problem. Second, regression approaches have been shown to produce reduced statistical errors  compared to rejection algorithms in a quite diverse range of statistical problems \cite{beaumont02,blumfrancois10,saulnier17}. Last, regression approaches are implemented in different ABC software including {\it DIYABC} \cite{cornuet14} and the R {\it abc} package \cite{csillery12}.

In this chapter, I introduce regression adjustment using a comprehensive framework that includes linear adjustment \cite{beaumont02} as well as more flexible adjustments such as non-linear models \cite{blumfrancois10}. The first section presents the main concepts underlying regression adjustment. The second section presents a theorem that compares theoretical properties of posterior distributions obtained with and without regression adjustment. The third section presents a practical application of regression adjustment in ABC. It shows that regression adjustment shrinks posterior distributions when compared to a standard rejection approach. The fourth section presents recent regression approaches for ABC that are not based on regression adjustment.

\section{Principle of regression adjustment}
\subsection {Partial posterior distribution}
Bayesian inference is based on the posterior distribution defined as
\begin{equation}
\label{eq:post}
\pi(\theta|y_{\rm obs}) \propto p(y_{\rm obs}|\theta)\pi(\theta)
\end {equation}
where $\theta \in \mathbb {R}^p$ is the vector of parameters, and $y_{\rm obs}$ 
are the data. Up to a renormalizing constant, the posterior distribution depends on the prior $\pi(\theta) $ and on the likelihood function $ p(y_{\rm obs}|\theta)$. In the context of ABC, inference is no longer based on the posterior distribution $\pi(\theta|y_{\rm obs})$ but on the {\it partial} posterior distribution $\pi(\theta|s_{\rm obs})$
  where $s_{\rm obs}$ is a $q$-dimensional vector of descriptive statistics. The partial posterior distribution is defined as follows
\begin {equation}
\label{eq:partialpost}
\pi (\theta |s _ {\rm obs}) \propto p(s_{\rm obs} | \theta) \pi (\theta).
\end{equation}
Obviously, the partial posterior is equal to the posterior if the descriptive statistics $s_{\rm obs}$ are sufficient for the parameter $\theta$.

\subsection {Rejection algorithm followed by adjustment}

To simulate a sample from the partial posterior $p(\theta| s _ {obs})$, the rejection algorithm followed by adjustment works as follows
\begin{VF}
\begin{enumerate}
\item Simulate $n$ values $\theta^{(i)}$, $i =1,\dots, n$, according to the prior distribution $\pi$.
\item Simulate descriptive statistics $s^{(i)} $ using the generative model $p(s^{(i)}|\theta^{(i)}) $.
\item Associate with each pair $ (\theta^{(i)},s^{(i)})$ a weight $ w^{(i)} \propto K_h(\| s^{(i)} - s_ {\rm obs}\|) $ where $ \| \cdot - \cdot \|$ is a distance function, $h>0$ is the bandwidth parameter, and $K$ is an univariate statistical kernel with $K_h(\| \cdot \|)=K(\| \cdot \|/h)$.
\item Fit a regression model where the response is $\theta$ and the predictive variables are the summary statistics $s$ (equations (\ref{eqn:reg}) or (\ref{eqn:reg2})). Use a regression model to adjust the $\theta^{(i)}$ in order to produce a weighted sample of adjusted values. Homoscedastic adjustment (equation (\ref{eqn:adj})) or heteroscedastic adjustment (equation (\ref{eqn:adjh})) can be used to produce a weighted sample $(\theta^{(i)}_c,w^{(i)})$, $i =1,\dots, n$, which approximates the posterior distribution.
\end{enumerate}
\end{VF}

To run the rejection algorithm followed by adjustment, there are several choices to make. The first choice concerns the kernel $K$. Usual choices for $K$ encompass uniform kernels that give a weight of 1 to all accepted simulations and zero otherwise \cite{pritchard99} or the Epanechnikov kernel for a smoother version of the rejection algorithm \cite{beaumont02}. However, as for traditional density estimation, the choice of statistical kernel has a weak impact on estimated distribution \cite{silverman86}. The second choice concerns the threshold parameter $h$. For kernels with a finite support, the threshold $h$ corresponds to (half) the window size within which simulations are accepted. For the theorem presented in section \ref{sec:theorem}, I assume that $h$ is chosen  without taking into account the simulations $s^{(1)},\dots,s^{(n)}$. This technical assumption does not hold in practice where we generally choose to accept a given percentage $p$, typically $1\%$ or $0.1\%$, of the simulations. This practice amounts at setting $h$ to the first $p$-quantile of the distances $\| s^{(i)} - s_ {\rm obs}\|$. A theorem where the threshold depends on simulations has been provided \cite{biau15}. Choice of threshold $h$ corresponds to bias-variance tradeoff. When choosing small values of $h$, the number of accepted simulations is small and estimators might have a large variance. By contrast, when choosing large values of $h$, the number of accepted simulations is large and estimators might be biased \cite{blum10}. 

\subsection{Regression adjustment}

The principle of regression adjustment is to adjust simulated parameters $\theta^{(i)}$ with nonzero weights $w^{(i)}>0$ in order to account for the difference between the simulated statistics $s^{(i)}$ and the observed one $s_{\rm obs}$. To adjust parameter values, a regression model is fitted in the neighborhood of $s_{\rm obs} $
\begin{equation}
\label{eqn:reg}
\theta^{(i)}=m(s^{(i)}) + \varepsilon, \quad i=1,\cdots,n
\end {equation}
where $m(s)$ is the conditional expectation of $\theta $ given $s$ and $\varepsilon $ is the residual. The regression model of equation (\ref{eqn:reg}) assumes {\it homoscedasticity}, i.e. it assumes that the variance of the residuals does not depend on $s$. To produce samples from the partial posterior distribution, the $\theta^{(i)}$'s are adjusted as follows
\begin{eqnarray}
\label{eqn:adj}
\theta^{(i)}_c & = & \hat {m} (s_{\rm obs}) + \hat {\varepsilon}^{(i)} \\ \nonumber
& = &  \hat{m} (s_{\rm obs}) + (\theta^{(i)} - \hat{m}(s^{(i)})),
\end{eqnarray}
where $\hat{m}$ represents an estimator of the conditional expectation of $\theta$ given $s$, and $\hat {\varepsilon}^{(i)}$ is the $i^{\rm th}$ empirical residual. In its original formulation, regression adjustment assumes that $m$ is a linear function \cite{beaumont02} and it was later extended to non-linear adjustments \cite{blumfrancois10}.

\begin{figure}
\includegraphics[width=350pt]{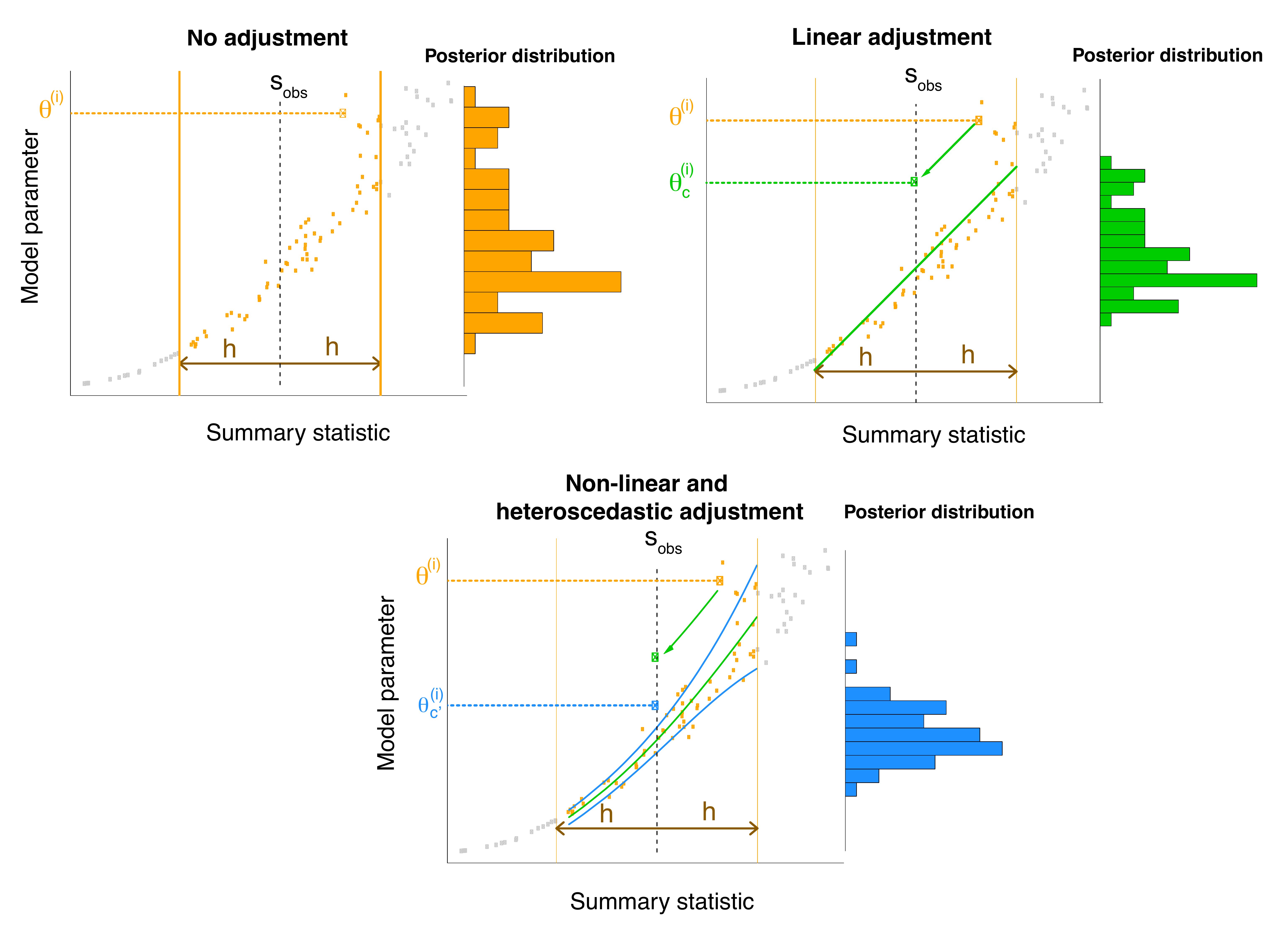}
\caption[Graphical description of regression adjustment.]{Posterior distributions obtained with and without regression adjustment. The shrinking effect of adjustment is visible since the posterior variance of the green and blue histograms is reduced compared to the posterior variance of the orange histogram.}
\label{fig:fig1}
\end{figure}

The homoscedastic assumption of equation (\ref{eqn:reg}) may not be always valid. When the number of simulations is not very large because of computational constraints, local approximations, such as the homoscedastic assumption, are no longer valid because the neighborhood corresponding to simulations for which $w^{(i)}\neq0$ is too large. Regression adjustment can account for heteroscedasticity that occurs when the variance of the residuals depend on the summary statistics. When accounting for heteroscedasticity, the regression equation can be written as follows \cite{blumfrancois10}

\begin{equation}
\label{eqn:reg2}
\theta^{(i)}=m(s^{(i)}) + \sigma(s^{(i)})\zeta, \quad i=1,\cdots,n
\end{equation}
where $\sigma({\bf s})$ is the square root of the conditional variance of $\theta$ given $s$, and $\zeta$ is the residual. Heteroscedastic adjustment involves an additional scaling step in addition to homoscedastic adjustment (\ref{eqn:adj}) (Figure \ref{fig:fig1})

\begin{eqnarray}
\label{eqn:adjh}
\theta^{(i)}_{c'}  & = & \hat{m} (s_{\rm obs}) +\hat{\sigma}(s_{\rm obs}) \hat{\zeta}^{(i)}\nonumber \\
&= & \hat{m}(s_{\rm obs}) + \frac{\hat{\sigma}(s_{\rm obs})}{\hat{\sigma}(s^{(i)})} (\theta^{(i)} - \hat{m}(s^{(i)})),
\end{eqnarray}
where $\hat{m}$ and $\hat{\sigma}$ are estimators of the conditional mean and of the conditional standard deviation.

\subsection{Fitting regression models}
Equations (\ref{eqn:adj}) and (\ref{eqn:adjh}) of regression adjustment depend on the estimator of the conditional mean $\hat{m}$ and possibly of the conditional variance $\hat{\sigma}$. Model fitting is performed using weighted least squares. The conditional mean is learned by minimizing the following  weighted least square criterion
\begin{equation}
\label{eqn:fit1}
E(m)=\sum_{i=1}^n (\theta^{(i)}-m(s^{(i)}))^2 w^{(i)}.
\end{equation}
For linear adjustment, we assume that $m(s)=\alpha+\beta s$ \cite{beaumont02}. The parameters $\alpha$ and $\beta$ are inferred by minimizing the weighted least square criterion given in equation (\ref{eqn:fit1}).

For heteroscedastic adjustment (equation (\ref{eqn:adj})), the conditional variance should also be inferred. The conditional variance is  learned after minimization of a least square criterion. It is obtained by fitting a regression model where the answer is the logarithm of the squared residuals. The weighted least squares criterion is given as follows
$$
E(\log{\sigma^2})=\sum_{i=1}^n \left(\log(({\hat {\varepsilon}}^{(i)})^2)-\log{\sigma^2(s)} \right)^2 w^{(i)}.
$$

Neural networks has been proposed to estimate $m$ and $\sigma^2$ \cite{blumfrancois10}. This choice was motivated by the possibility offered by neural networks to reduce the dimension of descriptive statistics via an internal projection on a space of lower dimension \cite{ripley94}.

In general, the assumptions of homoscedasticity and linearity (equation (\ref{eqn:reg})) are violated when the percentage of accepted simulation is large. By contrast, heteroscedastic and non-linear regression models (equation (\ref{eqn:reg2})) are more flexible. Because of this additional flexibility, the estimated posterior distributions obtained after heteroscedastic and non-linear adjustment is less sensitive to the percentage of accepted simulations \cite{blumfrancois10}. In a coalescent model where the objective was to estimate the mutation rate, heteroscedastic adjustment with neural networks was found to be less sensitive to the percentage of accepted simulations than linear and homoscedastic adjustment \cite{blumfrancois10}. In a model of phylodynamics, it was found again that statistical error obtained with neural networks decreases at first---because the regression method requires a large enough training dataset---and then reaches a plateau \cite{saulnier17}. However for larger phylodynamics dataset, statistical error obtained with neural networks increases for higher tolerance values. Poor regularization or the limited size of neural networks were advanced as putative explanations  \cite{saulnier17}.

In principle, estimation of the conditional mean $m$ and of the conditional variance $\sigma^2$ can be performed with different regression approaches. For instance, the \verb R  {\it abc} package implements different regression models for regression adjustment including linear regression, ridge regression and neural networks \cite{csillery12}. Lasso regression is another regression approach that can be considered. Regression adjustment based on lasso was shown to provide smaller errors than neural network in a phylodynamic model \cite{saulnier17}. An advantage of lasso, ridge regression and neural networks compared to standard multiple regression is that they account for the large dimension of the summary statistics using different regularization techniques. Instead of considering regularized regression, dimension reduction is an alternative where the initial summary statistics are replaced by a reduced set of summary statistics or a combination of the initial summary statistics \cite{fearnhead12,blum13}. The key and practical advantage of regression approaches with regularization is that they implicitly account for the large number of summary statistics and the additional step of variable selection can be avoided.

\subsection{Parameter transformations}
When the parameters are bounded or positive, parameters can be transformed before regression adjustment. Transformations guarantee that the adjusted parameter values lie in the range of the prior distribution \cite{beaumont02}. An additional advantage of the ${\it log}$ and ${\it logit}$ transformations is that they stabilize the variance of the regression model and make regression model (\ref{eqn:reg}) more homoscedastic  \cite{blum10}.

Positive parameters are regressed on a logarithm scale $\phi=\log(\theta)$,
$$
\phi=m({\bf s}) + \varepsilon.
$$
Parameters are then adjusted on the logarithm scale
$$
\phi^{(i)}_c=\hat {m} (s_{\rm obs})  + (\phi^{(i)} - \hat{m}(s^{(i)})).
$$
The final adjusted values are obtained by exponentiation of the adjusted parameter values
$$
\theta^{(i)}_c = \exp(\phi^{(i)}_c). 
$$
Instead of using a logarithm transformation, bounded parameters are adjusted using a logit transformation. Heteroscedastic adjustment can also be performed after log or logit transformations.

\subsection{Shrinkage}

An important property of regression adjustment concerns posterior shrinkage. When considering linear regression, the empirical variance of the residuals is smaller than the total variance. In addition, residuals are centered for linear regression. These two properties imply that for linear adjustment, the empirical variance of $\theta^{(i)}_c$ is smaller than the empirical variance of the non-adjusted values $\theta^{(i)}$ obtained with the rejection algorithm. Following homoscedastic and linear adjustment, the posterior variance is consequently reduced. For non-linear adjustment, shrinkage property has also been reported  and the additional step generated by heteroscedastic adjustment does not necessarily involve additional shrinkage when comparing  $\theta^{(i)}_c$ to $\theta^{(i)}_{c'}$ \cite{blum10}.

\section{Theoretical results about regression adjustment}\label{sec:theorem}
The following theoretical section is technical and can be skipped by readers not interested by mathematical results about ABC estimators based on regression adjustment. In this section, we give the main theorem that describes the statistical properties of posterior distributions obtained with or without regression adjustment. To this end, the estimators of the posterior distribution are defined as follows
\begin{equation}
\label{eqn:estim12}
\hat{\pi}_j(\theta| s_{\rm obs})=\sum_{i=1}^n \tilde{K}_{h'}(\theta^{(i)}_j-\theta) w^{(i)}, \; j=0,1,2,
\end{equation}
where $\theta^{(i)}_0=\theta^{(i)}$ (no adjustment), $\theta^{(i)}_j=\theta^{(i)}_c$ for $j=1,2$ (homoscedastic adjustment), $\tilde{K}$ is an univariate kernel, and $\tilde{K}_{h'}(\cdot)=\tilde{K}(\cdot)/h'$. Linear adjustment corresponds to $j=1$ and quadratic adjustment corresponds to $j=2$.
In non-parametric statistics, estimators of the conditional density with adjustment have already been proposed \cite{hyndman96,hansen04}.

To present the main theorem, we introduce the following notations: if $X_n$ is a sequence of random variables and $a_n$ is a deterministic sequence, the notation $X_n= o_P (a_n) $ means that $ X_n / a_n$ converges to zero in probability and $X_n=O_P (a_n)$ means that the ratio $X_n/a_n $ is bounded in probability when $n$ goes to infinity. The technical assumptions of the theorem are given in the appendix of \cite{blum10}.

\begin{theorem}
\label{th:reject_beaumont1}
We assume that conditions (A1)-(A5) given in the appendix of \cite{blum10} hold. The bias and variance of the estimators $\hat{\pi}_j(\theta| s_{\rm obs})$, $j=0,1,2$ are given by
\begin{equation}
\label{eqn:bias_rej_beaumont}
E[\hat{\pi}_j(\theta| s_{\rm obs})-\pi(\theta | s_{\rm obs})]  =  C_1{h'}^2+C_{2,j}h^2 + O_P((h^2+h'^2)^2)+O_P(\frac{1}{nh^q}),  
\end{equation}

\begin{equation}
\label{eqn:var_rej_beaumont}
{\rm Var}[\hat{\pi}_j(\theta | s_{\rm obs})]=\frac{C_3}{nh^qh'} (1+o_P(1)),
\end{equation}
where $q$ is the dimension of the vector of summary statistics and the constants $C_1$, $C_{2,j}$ et $C_3$ are given in \cite{blum10}.
\end{theorem}

\noindent Proof: See \cite{blum10}.

There are other theorems that provide  asymptotic biases and variances of ABC estimators but they do not study the properties of  estimators arising after regression adjustment. Considering posterior expectation (e.g. posterior moments) instead of the posterior density, \cite{barber15} provides asymptotic bias and variance of an estimator obtained with rejection algorithm. \cite{biau15} studied asymptotic properties when the window size $h$ depends on the data instead of being fixed in advance. Another version of ABC called {\it lazy ABC} exists and provides a bias proportional to $h$ instead of $h^2$ while keeping the same variance of  the order of $1/(nh^q)$, which is inversely proportional to the acceptance probability in the rejection algorithm \cite{fearnhead12}. 
{\par{\bf Remark 1. Curse of dimensionality}}
The mean square error  of the estimators is the sum of the squared bias and of the variance. With elementary algebra, we can show that for the three estimators $\hat{\pi}_j(\theta | s_{\rm obs})$, $j=0,1,2$, the mean square error is of the order of $n^{-1/(q+5)}$ for an optimal choice of $h$. The speed with which the error approaches 0 therefore decreases drastically when the dimension of the descriptive statistics increases. This theorem highlights (in an admittedly complicated manner) the importance of reducing the dimension of the statistics. However, the findings from these asymptotic theorems, which are classic in non-parametric statistics, are often much more pessimistic than the results observed in practice. It is especially true because asymptotic theorems in the vein of Theorem \ref{th:reject_beaumont1} do not take into account correlations between summary  statistics \cite{scott09}.
{\par {\bf Remark 2. Comparing biases of estimators with and without adjustment }}
It is not possible to compare biases (i.e. the constant $C_{ 2,j}$, $ j = 0,1,2 $) for any statistical model. However, if we assume that the residual distribution of $ \varepsilon $ in equation (\ref{eqn:reg}) does not depend on $s$, then the constant $C_{ 2,2}$ is 0. When assuming homoscedasticity, the estimator that achieves asymptotically the smallest mean square error is the estimator with quadratic adjustment $\hat{p}_2(\theta| s_{\rm obs})$. Assuming additionally that the conditional expectation $m$ is linear in $s$, then both $ \hat{p}_1(\theta | s_{\rm obs})$ and $\hat{p}_2(\theta | s_{\rm obs})$ have a mean square error lower than the error obtained without adjustment.
%
%
%
\section{Application of regression adjustment to estimate admixture proportions using polymorphism data}\label{sec:theorem}

\begin{figure}
\includegraphics[width=200pt]{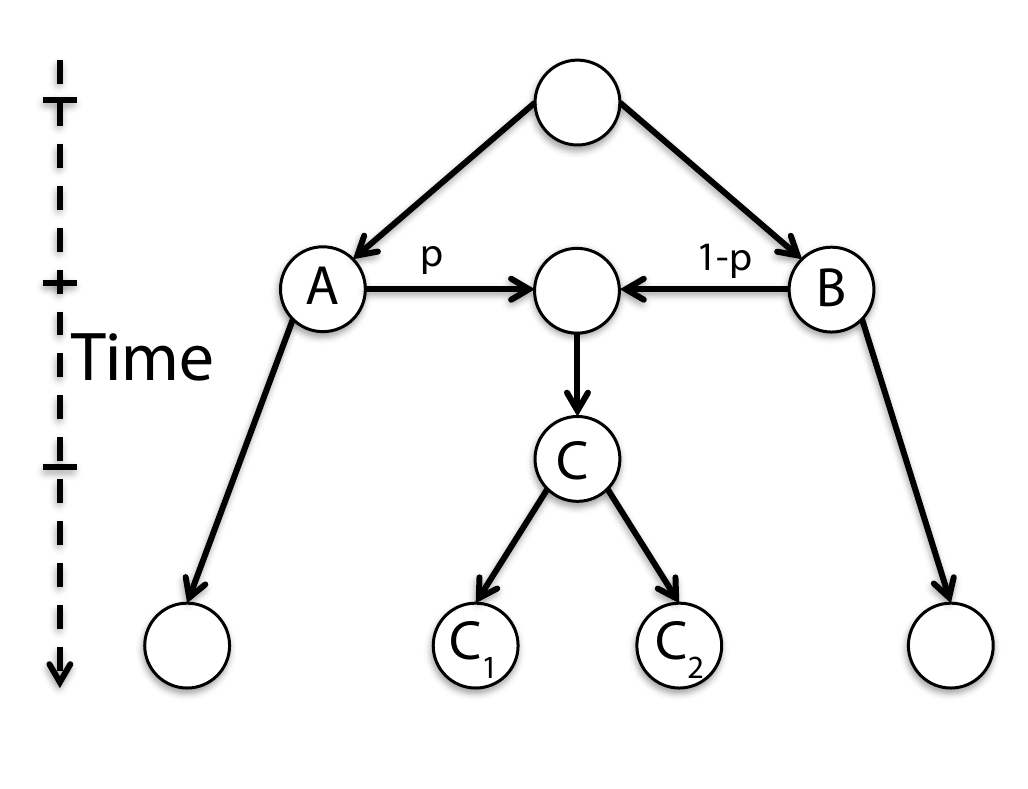}
\caption[Graphical description of the model of admixture between sister species.]{Graphical description of the model of admixture between sister species.  Two populations ($A$ and $B$) diverge in the past and admixed with admixture proportions $p$ and $1-p$ to form a new hybrid species $C$ that subsequently diverge to form two sister species ($C_1$ and $C_2$).}
\label{fig:fig2}
\end{figure}

To illustrate regression adjustment, I consider an example of parameter inference in population genetics. Description of coalescent modeling in population genetics is out of the scope of this chapter and we refer interested readers to dedicated reviews \cite{hudson90,rosenberg02}. This example illustrates that ABC can be used to infer evolutionary events such as admixture between sister species. I assume that two populations ($A$ and $B$) diverged in the past and admixed with admixture proportions $p$ and $1-p$ to form a new hybrid species $C$ that subsequently split to form two sister species $C_1$ and $C_2$ (Figure \ref{fig:fig2}). Simulations are performed using the software DIYABC \cite{cornuet14}. The model of Figure \ref{fig:fig2} corresponds to a model of divergence and admixture between species of a complex of species from the butterfly gender {\it Coenonympha}. We assume that 2 populations of the Darwin's Heath ({\it Coenonympha darwiniana}) originated through hybridization between the Pearly Heath ({\it Coenonympha arcania}) and the Alpine Heath ({\it Coenonympha gardetta}) \cite{capblancq15}. A total of 16 summary statistics  based on Single Nucleotide Polymorphisms (SNPs) are used for parameter inference \cite{capblancq15}. A total of $10^6$ simulations are performed and the percentage of accepted simulations is of $0.5\%$. 

\begin{figure}
\includegraphics[width=200pt]{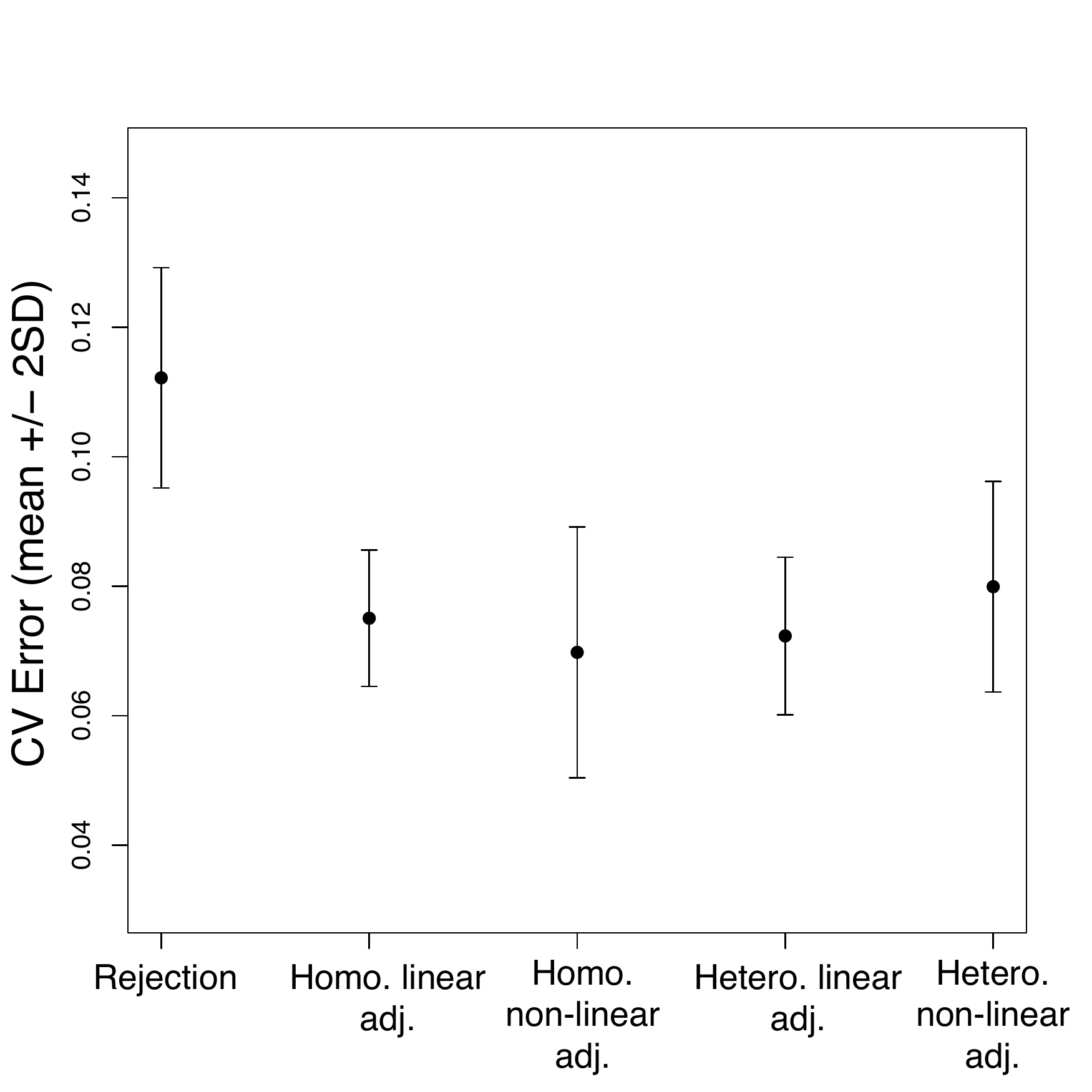}
\caption[Cross-validation error.]{Errors obtained when estimating the admixture proportion $p$ with different ABC estimators. The errors are obtained with cross-validation and error bars (two standard deviations) are estimated with bootstrap. {\it adj.} stands for adjustment.}
\label{fig:fig3}
\end{figure}

I consider four different forms of regression adjustment: linear and homoscedastic adjustment, non-linear (neural networks) and homoscedastic adjustment, linear and heteroscedastic adjustment, non-linear and heteroscedastic adjustment. All adjustments were performed with the \verb R  package {\it abc} \cite{csillery12,R}. I evaluate parameter inference using a cross-validation criterion \cite{csillery12}. The cross-validation error decreases when considering linear adjustment (Figure \ref{fig:fig3}). However, considering heteroscedastic instead of homoscedastic adjustment does not provide an additional decrease of the cross-validation error (Figure \ref{fig:fig3}).

\begin{figure}
\includegraphics[width=200pt]{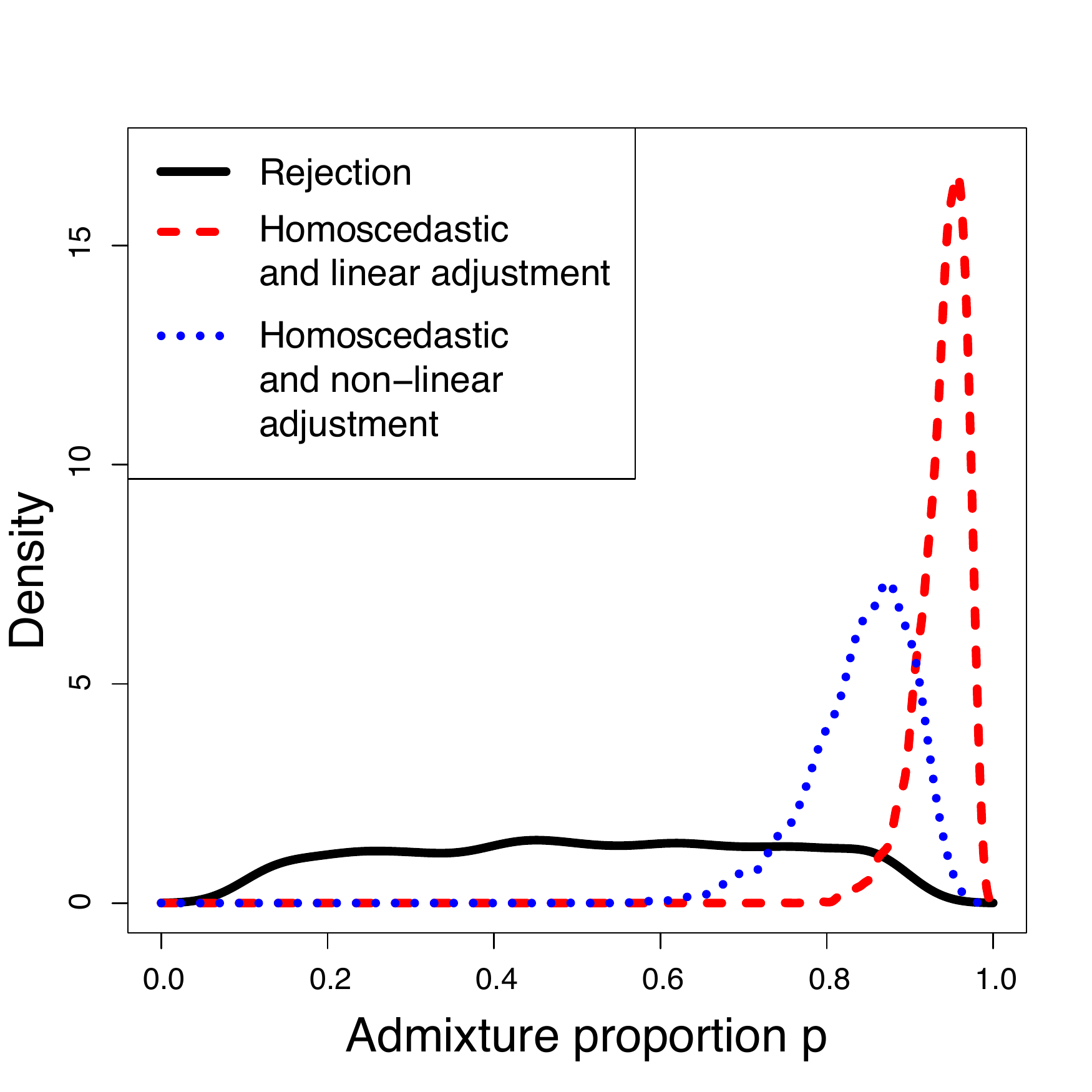}
\caption[Posterior distribution of $p$.]{Posterior distribution of the admixture coefficient $p$ obtained using real data from a butterfly species complex to compute observed summary statistics. This example shows that regression adjustment not only shrinks credibility intervals but can also shift posterior estimates. The admixture coefficient $p$ measures the relative contribution of {\it Coenonympha arcania} to the ancestral {\it C. darwiniana} population.}
\label{fig:fig4}
\end{figure}

Then, using real data from a butterfly species complex, we compare the posterior distribution of the admixture proportion $p$ obtained without adjustment, with linear and homoscedastic adjustment, and with non-linear and homoscedastic adjustment (Figure \ref{fig:fig4}). For this example, considering regression adjustment considerably changes the shape of the posterior distribution. The posterior mean without adjustment is of $p=0.51$ ($95\%\, C.I.= (0.12,0.88)$). By contrast, when considering linear and homoscedastic adjustment, the posterior mean is of $0.93$ ($95\% \,C.I.= (0.86,0.98)$. When considering non-linear and homoscedastic adjustment, the posterior mean is $0.84$ ($95\% \,C.I.= (0.69,0.93)$. Regression adjustment confirms a larger contribution of {\it C. arcania} to the genetic composition of the ancestral {\it C. darwiniana} population \cite{capblancq15}. This example shows that regression adjustment not only shrinks credibility intervals but can also shift posterior estimates. Compared to rejection, the posterior shift observed with regression adjustments provides a result that is more consistent with published results \cite{capblancq15}.

\section{Regression methods besides regression adjustment}
There are other regression methods besides regression adjustment that have been proposed to estimate $E[\theta|s_{obs}]$ and $\pi(\theta| s _ {obs})$ using the simulations as a training set. A first set of methods consider kernel methods to perform regression \cite{nakagome13}. The principle is to define a kernel function $K$ to compare observed statistics to simulated summary statistics. Because of the so-called kernel trick, regression with kernel methods amounts at regressing $\theta$ with $\Phi(s)$ where $<\Phi(s),\Phi(s')>=K(s,s')$ for two vector of summary statistics $s$ and $s'$. Then, an estimate of the posterior mean is obtained as follows
\begin{equation}
\label{eq:kernel}
E[\theta | s _ {obs}] =\sum_{i=1}^n w_i\theta^{(i)},
\end{equation}
where $w_i$ depends on the inverse the Gram matrix containing the values $K(s^{(i)},s^{(j)})$ for $i,j=1\dots,n$. A formula to estimate posterior density can also be obtained in the same lines as formula (\ref{eq:kernel}). Simulations suggest that kernel ABC gives better performance than regression adjustment when high-dimensional summary statistics are used. For a given statistical error, it was reported that fewer simulations should be performed when using kernel ABC  instead of regression adjustment \cite{nakagome13}. Other  kernel approaches have been proposed for ABC where simulated and observed samples or summary statistics are directly compared through a distance measure between  empirical probability distributions  \cite{mitrovic16,park16}.

Another regression method in ABC that does not use regression adjustment considers quantile regression forest \cite{marin16}.  Generally used to estimate conditional mean, random forests also provide information about the full conditional distribution of the response variable \cite{meinshausen06}. By inverting the estimated conditional cumulative distribution function of the response variable, quantiles can be inferred \cite{meinshausen06}. The principle of quantile regression forest is to use random forests in order to give a weight $w_i$ to each simulation $(\theta^{(i)},s^{(i)})$. These weights  are then used to estimate the conditional cumulative posterior distribution function $F(\theta|s_{obs})$ and to provide posterior quantiles by inversion. An advantage of quantile regression forest is that tolerance rate should not be specified and standard parameters of random forest can be considered instead. A simulation study of coalescent models shows that regression adjustment can shrink posterior excessively by contrast to quantile regression forest \cite{marin16}.

\section{Conclusion}
This chapter introduces regression adjustment for Approximate Bayesian Computation \cite{beaumont02,blumfrancois10}. We explain why regression adjustment  shrinks posterior distribution which is a desirable feature because credibility intervals obtained with rejection methods can be too wide \cite{blum10}. When inferring admixture with SNP data in a complex of butterfly species, the posterior distribution obtained with regression adjustment was not only shrunk when compared to standard rejection but also shifted to larger values, which confirm results obtained for this species complex with other statistical approaches \cite{capblancq15}. We have introduced different variants of regression adjustment and it might be difficult for ABC users to choose which adjustment is appropriate in their context. We argue that there is no best strategy in general. In the admixture example, we found, based on a cross-validation error criterion, that homoscedastic linear adjustment provides considerable improvement compared to rejection. More advanced adjustments provide almost negligible improvements if no improvement at all. However, in a model of phylodynamics, non-linear adjustment was reported to achieve considerable improvement compared to linear adjustment  \cite{saulnier17}. In practical applications of ABC, we suggest to compute errors such as cross-validation estimation errors to choose a particular method for regression adjustment.

With the rapid development of complex machine learning approaches, we envision that regression approaches for Approximate Bayesian Computation can be further improved to provide more reliable inference for complex models in biology and ecology.

\bibliographystyle{plain}
\bibliography{bib_MB.bib}

\end{document}